\documentstyle[epsfig,12pt]{article}

\begin{document}

\title{Scattering Theory of Mesoscopic Detectors}
\author{M. B\"uttiker and S. Pilgram\\
Dept. Phys. Th\'eorique, Universit\'e de Gen\`eve,\\
24, quai Ernest-Ansermet, 1211 Gen\`eve 4, Switzerland}

\maketitle

\begin{abstract}
We consider a two-level system coupled to a mesoscopic
two-terminal conductor that acts as measuring device.
As a convenient
description of the conductor we introduce its
scattering matrix. We show how its elements can be used to
calculate the relaxation and decoherence
rates of the two-level system.
Special emphasis is laid on the charge screening in
the conductor that becomes important in the
many-channel limit.
Finally we give some examples that illustrate charge
screening in different limits.
\end{abstract}



\begin{figure}[htb]
\begin{center}
\leavevmode


\psfig{file=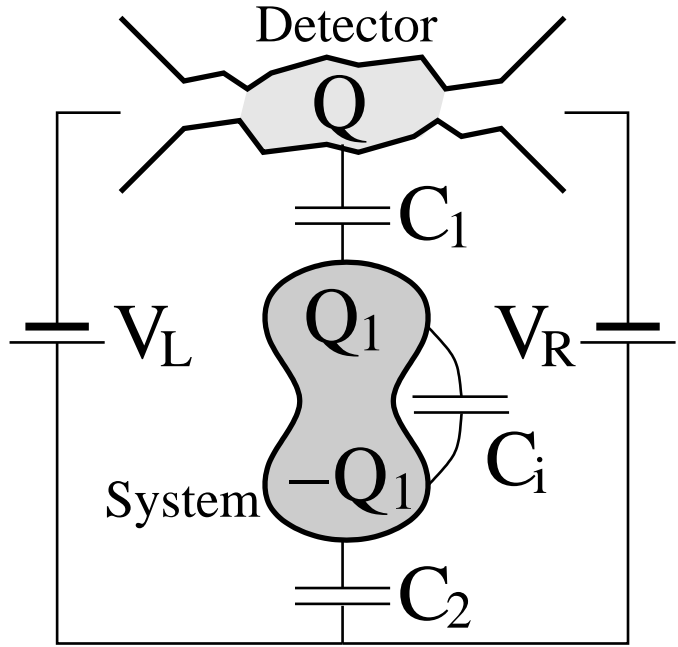,width=9cm}

\vspace{5mm}

\caption{A mesoscopic detector is capacitively coupled
to one side of a double dot.
}
\label{Sketch1}
\end{center}
\end{figure}

The detection of quantum states by means of mesoscopic
detectors is of fundamental importance from the point of
view of quantum measurement theory. It is also 
important due to the recent interest
in different schemes of quantum computation for which the read-out
process must be a key step \cite{exp}. A detector, typically, has a back
action \cite{averin} on the measured system and thus can also be used
to introduce decoherence in a controlled way.
Here we are concerned with a simple
prototype of a detector that consists of a mesoscopic conductor
capacitively coupled to the state of a nearby quantum system
(see Fig. (\ref{Sketch1})). The most basic
quantum system is a two state system
here taken to be two mesoscopic quantum dots
weakly coupled to each other.

Recent experiments demonstrated elegantly the effect of a
detector on a phase coherent mesoscopic system
\cite{Buks1,Sprinzak1,Field1,Smith1}.
Different aspects of weak measurement were addressed
in several theoretical discussions: the relation between
detector noise and decoherence \cite{Levinson1,Levit1},
their connection to measurement time \cite{Aleiner1}
and to scattering theory \cite{Buks1,Stodolsky1,MBMartin1}.
Master equation approaches have been used to study the
time evolution of system and detector
\cite{Gurvitz1,Schoen1}. These equations
have been refined to describe also conditional
evolution depending on the outcome of the measurement
\cite{Korotkov1,Milburn1}. Tunnel contacts and
single electron transistors have been identified
as candidates for efficient
measurement devices \cite{Averin2}.

For such devices it is natural to ask whether their
speed can be improved by increasing their size. One is tempted
to expect that the sensitivity of a detector grows when
we add more and more conductance channels. The bigger the detector
the less the uncertainty due to the shot noise compared to the signal
of the detector. However, this effect competes with the
charge screening between different conductance channels
through the detector that grows as well with increasing
channel number. This screening tends to suppress the sensitivity
of the detector.

Charge screening is often neglected in mesoscopic physics.
Typically investigated quantities are dc-currents and low frequency
current noises which are insensitive to displacement currents in the
system. However, there are quantities that depend
crucially on screening: ac-conductance measurements pick up
screening currents in the system:
At contacts which permit the exchange of particles these become noticeable
only if the frequency gets of the order
of the inverse dwell time in the system;
at nearby capacitors only displacement currents contribute
and can be measured at arbitrary low frequencies \cite{Buettiker2}.
Furthermore non-linear current-voltage characteristics
depend on the charge distribution inside mesoscopic conductors
\cite{Christen}.
In the case of mesoscopic detectors the quantity of interest
is the charge fluctuations in the detector. Like the currents
induced into a nearby gate this quantity is subject to screening
already in the low-frequency limit! Earlier work focused on the effect of
screening on the decoherence rate induced by the measurement
process under the assumptions that both detector and system
are described by a scattering matrix \cite{MBMartin1,Ankara}.
It is the aim of this work to investigate the effect
of charge screening in the detector on the DD system
shown in Fig. \ref{Sketch1}.

\section{The model}

The model we consider is shown in Fig. \ref{Sketch1}.
A double dot (DD) plays the role of a two-level system:
The topmost electron in the DD can either occupy the
upper or the lower dot.
A mesoscopic two-terminal conductor (MC) serves as
a detector: its conductance is sensitive to the charge on
the upper dot.
A set of capacitances $C_1,C_2,C_i$ describes the coupling
between charges of system $Q_1$ and detector $Q$
(we often use the capacitance in series
$C^{-1} = C_1^{-1} + C_2^{-1} + C_i^{-1}$).
Single electron movement in such a setup was recently
measured \cite{Smith1}.

Three different time scales describe the interaction of system
and detector: In the system  we distinguish the thermal
{\it relaxation} to an equilibrium distribution
(described by a rate $\Gamma_{rel}$) and the
often much faster {\it decoherence} of
superpositions of states in the upper and lower dot
(described by a rate $\Gamma_{dec}$). The decoherence
depends as well on temperature $kT$ as on the
voltage difference $e|V|$ between the terminals of the detector.
Applying such a voltage difference permits to
measure the state of the system which leads to additional
decoherence. The {\it measurement} process takes some
time which is needed to overcome the uncertainty
due to shot noise in the
detector and is characterized by a third rate
$\Gamma_m$. The decoherence rate $\Gamma_{dec}$ at zero temperature
is intimately related to the measurement rate $\Gamma_m$ and satisfies
the inequality $\Gamma_{dec} \ge \Gamma_m$ \cite{Schoen1,Averin1}.
An important measure of the quality of the detector is the
efficiency $\eta = \Gamma_m /\Gamma_{dec}$. Ideally one would
like to find detectors with an efficiency $1$.

Even in the single channel case an ideal detector $\eta = 1$
is found only under some
special conditions (time-reversal and space-inversion symmetry
\cite{Averin1}). It is shown in Ref. \cite{Pilgram1}
that an efficient multi-channel detector is subject to a {\it third} condition
that links sensitivities and shot noises of different channels
$(dT_n / dU)/(R_n T_n) = \mbox{const.}$
where $T_n = 1-R_n$ denotes the transmission probability of channel $n$.
A violation of this condition can reduce
the efficiency of multichannel detectors $\eta$
drastically in the presence of disorder. In general a large
mesoscopic detector is therefore not much faster than a single
channel detector.

In this publication we concentrate on another property that lowers
the detection speed of multichannel detectors.
The charge fluctuation in one channel can be screened by the charge
of other channels.
This effect will depend strongly on
the geometry of the detector.
In order to be able to describe a
large variety of detectors we therefore represent the MC by a scattering
matrix
$s_{\alpha\beta}$ that connects in- and outgoing states ($\alpha,\beta$
label left and right reservoir). This enables us to treat
multi-channel MCs with arbitrary transmission probabilities $T_n$.
We only consider the case of weak coupling between MC and DD and
may thus use a standard master
equation (Bloch-Redfield approach \cite{Carmichael1}) in lowest order
perturbation theory to study the evolution of the reduced density
matrix of the DD. On this level of approximation the dynamics of the
DD is influenced only via the charge fluctuation spectrum $S_{QQ}$
of the MC. At low frequencies this spectrum is fully characterized by
a generalized Wigner-Smith time delay matrix \cite{Pedersen1}
($\beta\gamma$ label the reservoirs)
\begin{equation}
\label{Density Elements}
N_{\beta\gamma} = \frac{1}{2\pi i} \sum_{\alpha}
s_{\beta\alpha}^{\dagger} \frac{ds_{\gamma\alpha}}{dU}.
\end{equation}
The derivative $d/dU$ is taken with respect to the
electrostatic potential in the MC.
The following constants
that appear also in the context of ac-transport
\cite{Buettiker2} describe the geometry
of the detector ($e$ denotes the electron charge)
\begin{equation}
\label{Four Parameters}
\begin{array}{ccc}
D = e^2 \mbox{Tr} N, & &
C_{\mu}^{-1} = C^{-1} + D^{-1},\\
\quad\\
R_q = \frac{1}{2}\frac{\left(\mbox{Tr} N^2\right)}{\left(\mbox{Tr}
N\right)^2}, &
R_v = \frac{\left(\mbox{Tr} N_{12}N_{21}\right)}{\left(\mbox{Tr}
N\right)^2}, &
R_m = \frac{1}{4\pi^2} 
\frac{\left(\sum \frac{dT_n}{dU}\right)^2}
{\left(\mbox{Tr}
N\right)^2
 \left(\sum R_n T_n \right)}
.\\
\end{array}
\end{equation}
$D$ corresponds
to the density of states at Fermi energy in the scattering region,
$C_{\mu}$ is an effective electrochemical capacitance that
characterizes the strength of interaction, $R_q$ expresses the
equilibrium contribution \cite{Buettiker2}
to the charge fluctuation spectrum $S_{QQ}$,
and $R_v$ the non-equilibrium contribution \cite{MBMartin1,Pedersen1}.
The fifth constant $R_m$ is of different origin \cite{Aleiner1}
and cannot be expressed by matrix (\ref{Density Elements}) only.
It describes the ratio of detector sensitivity and shot noise
in the presence of screening.

The two-level system
is conventionally represented by the Hamiltonian
$\hat{H}_{DD} = \frac{\epsilon}{2}\hat{\sigma}_z
 + \frac{\Delta}{2} \hat{\sigma}_x$ where
$\hat{\sigma}_i$ denote Pauli matrices.
The energy difference between upper and lower dot is $\epsilon$ and
$\Delta$
accounts for tunneling between the dots. The full level splitting is
thus $\Omega=\sqrt{\epsilon^2+\Delta^2}$.

\section{Relaxation and decoherence rates}

The relaxation, decoherence and measurement rates are given by the following
expressions \cite{Pilgram1}
(here and in the following we set $\hbar \equiv 1$):
\begin{equation}
\label{Central Result Relaxation}
\Gamma_{rel} =  2\pi \frac{\Delta^2}{\Omega^2}
\left(\frac{C_{\mu}}{C_i}\right)^2
R_q \frac{\Omega}{2}\coth \frac{\Omega}{2kT},
\end{equation}
\begin{equation}
\label{Central Result Decoherence}
\Gamma_{dec} = 
 2\pi \frac{\epsilon^2}{\Omega^2}
\left(\frac{C_{\mu}}{C_i}\right)^2
\left(R_q kT + R_v e|V|\right) + \Gamma_{rel}/2, 
\end{equation}
\begin{equation}
\label{Central Result Measurement}
\Gamma_m = 2\pi \left(\frac{C_{\mu}}{C_i}\right)^2
R_m e|V|.
\end{equation}
They have formally
the same appearance as the rates given in \cite{Schoen1}.
The geometric structure of the detector
is contained in the five parameters given in (\ref{Four Parameters}).
In the following we analyze the behavior of these parameters
to discuss the screening properties of the detector.

The parameter $R_q$ does not
exceed the range $1/2 > R_q > 1/2N$ where $N$ is the
dimension of the scattering matrix. This demonstrates
that the relaxation and decoherence rates
$\Gamma_{rel},\Gamma_{dec}$ do not simply scale with
the number of channels through the system.
The multichannel result
for the relaxation and decoherence rates cannot be
obtained as a sum of rates
due to each channel.
The constants $R_q$ and $R_v$ are renormalized
by a denominator $(\mbox{Tr} N)^2$ depending on all channels.
This denominator originates from screening.
For a large number $N$ of open
channels $R_q$
behaves as $1/N$ whereas the electrochemical
capacitance $C_{\mu}\rightarrow C$ tends to a constant.
The constant $R_v$ decreases also like $1/N$.
We find therefore the somewhat surprising result
that relaxation and decoherence decrease in the large
channel limit. This result is a consequence of
screening in the MC which reduces the charge fluctuations
with increasing channel number $N$.

\section{Coulomb interaction and screening}

We briefly explain the derivation of
our results.
The Coulomb Hamiltonian of our model contains three terms
\begin{equation}
\hat{H}_C = \frac{(\hat{Q}_1-\bar{Q}_0)^2}{2C_i} +
\frac{\hat{Q}_1\hat{Q}}{C_i}
          + \frac{\hat{Q}^2}{2C}.
\end{equation}
Its first term contributes to the level splitting
of the DD ($\bar{Q}_0$ is a background charge depending on the
applied voltage $(V_L+V_R)/2$). The charging energy $e^2/2C_i$
must be large compared to $kT,e|V|$ to allow us to
consider only two levels of the DD. The second term
$\hat{Q}_1\hat{Q}/C_i$ couples system and detector.
To derive a master equation for the reduced density
matrix we assume weak coupling and treat this term
perturbatively.
We apply a Markov approximation which is strictly speaking
only valid at long time scales (compared to the correlation time
of the detector). This permits us to consider only the
low-frequency matrix elements of the
charge operator that are given by the Wigner-Smith
matrix (\ref{Density Elements}).
The third term affects the
fluctuation spectrum $S_{QQ} = \int_{-\infty}^{+\infty} dt
\mbox{Re} \langle \hat{Q}(t) \hat{Q}(0) \rangle e^{i\omega t}$
of the charge operator $\hat{Q}$.
In contrast to earlier work (with the exception of Ref.
\cite{MBMartin1}) we do not
completely disregard this term, but include it
on the level of RPA. This allows us to treat screening
in a Gaussian approximation valid for 
geometries in which Coulomb blockade
effects are weak.

To get insight into the fluctuation spectrum, we study the
time evolution of the charge $\hat{Q}$ on the MC within RPA.
This charge is composed of
two components: the bare charge $\hat{Q}_b$ that arises from the
noninteracting problem and the screening charge which is
a response to the self-consistent potential $\hat{U}$
on the MC. We therefore have in frequency representation
\begin{equation}
\label{Charge RPA}
\hat{Q} = \hat{Q}_b - D(\omega) \hat{U},
\qquad \hat{Q} = C \hat{U} - \frac{C}{C_i} \hat{Q}_1.
\end{equation}
In the first equation, $D=-\partial \langle \hat{Q} \rangle / \partial U$
is the linear response function. The second
equation expresses the self-consistency condition for
the potential $\hat{U}$. These two equations can be combined
by eliminating the potential $\hat{U}$. The total charge
on the MC is found to be
\begin{equation}
\label{Total Charge}
\hat{Q} = \left(1+D(\omega)/C\right)^{-1}\left(\hat{Q}_b -
\frac{C}{C_i}\hat{Q}_1\right).
\end{equation}
This result shows nicely the two possible effects of the
random phase approximation. On the one hand, the prefactor
$\left(1+D(\omega)/C\right)^{-1}$ reduces the fluctuations
of the bare charge $\hat{Q}_b$ which is compensated partially
by a screening charge. On the other hand, we get a back action
of the charge $\hat{Q}_1$
of the upper dot on the charge $\hat{Q}$ in the MC.
However, this back action is small by a factor $C/C_i$
and can be omitted in second order perturbation
theory.

\section{Examples}

The examples of this section are not meant to
model the geometry of a realistic measuring apparatus.
But they demonstrate nicely the influence of the
geometry on the speed and efficiency of the
measuring process.

As a first generic example we discuss a
{\bf short quantum point contact}
which is defined by two sharp barriers
of distance $\ell$. Its
potential is given by $V(x,y,z) = Z(z) +
Y(x,y)$ with
\begin{equation}
Z(z) = (1/m) [v\delta(z+\ell/2) + w\delta(z-\ell/2)].
\end{equation}
Note that this potential violates inversion symmetry
for $v\neq w$. For convenience we introduce
the following parameters: the
wave vector $k_F = \sqrt{2mE_F}$
where $m$ is the effective mass of the
electron and $E_F$ the Fermi energy. 
We consider the limit of a MC
much shorter than the Fermi wave length
$\ell \ll \lambda_F$.
Using the general formula (\ref{Density Elements})
we calculate the density of state
matrix (we set $\hbar \equiv 1$)
\begin{equation}
\begin{array}{c}
N = \\
\quad\\
\frac{m\ell}{2\pi k_F}
\frac{1}{k_F^2 + (v+w)^2}
\left(\begin{array}{cc}
k_F^2 + 2w^2    & 2vw + ik_F(v-w)\\
2vw - ik_F(v-w) & k_F^2 + 2v^2
\end{array}\right).
\end{array}
\end{equation}
It is now straightforward to obtain
the measurement efficiency of the
MC at zero temperature
\begin{equation}
\eta = \frac{\Gamma_m}{\Gamma_{dec}} =\frac{R_m}{R_v} =
\frac{v^2w^2}{v^2w^2 + k_F^2(v-w)^2/4} < 1.
\end{equation}
For an asymmetric MC with $v\neq w$ the
efficiency is suboptimal.

\begin{figure}[htb]
\begin{center}
\leavevmode
\psfig{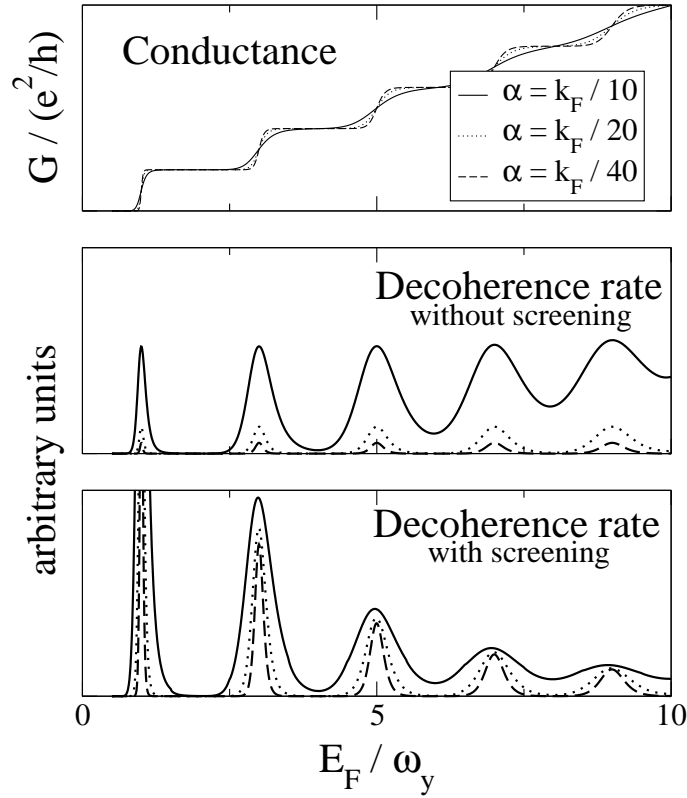}
\caption{
The effect of screening on the decoherence
rate $\Gamma_{dec}$ at zero temperature $kT$
and finite voltage bias $e|V|$ (see text).
}
\label{Integrable QPC}
\end{center}
\end{figure}

For a {\bf long quantum point contact}
with $\lambda_F \ll \ell$
charge screening gets effective.
The constants $R_q$ and $R_v$
that determine the relaxation and
decoherence rates then decrease with
increasing number of open
channels $N$. We illustrate this effect
using the following toy potential
\begin{equation}
Z(z) = \frac{V_{0}}{\cosh^2\alpha z}.
\end{equation}
Its scattering matrix can be found in \cite{Landau1}.
To get the full density of state matrix (\ref{Density Elements})
we replace the potential derivative $d/dU$ in
(\ref{Density Elements}) by a parametric derivative
$d/dV_{0}$.
The elements of (\ref{Density Elements})
can be expressed by digamma-functions.
As a confining potential we choose $Y = m \omega_y^2 y^2 / 2$.
As in a two-dimensional electron gas only the lowest mode
in z-direction is occupied. Fig. \ref{Integrable QPC}
compares the cases of ineffective charge screening $D \ll C$ and
effective screening $D \gg C$. It
shows the decoherence (measurement) rate at zero temperature $kT$ and
finite
bias $e|V|$ depending on the number of
open channels through the constriction. We change this number
by varying the confinement frequency $\omega_y$.
The decoherence rate reaches maximal values if one transmission channel
$n$
is half-open $T_n=1/2$.
This corresponds to a maximum in the sensitivity $dT_n/dV_{0}$.
In the case of ineffective screening the decoherence rate
does not depend on the number of open channels, whereas
screening reduces decoherence in the case of many open channels.

\section{Conclusions}

In this work we have investigated the weak measurement of
a two state system (a double quantum dot) capacitively
coupled to a mesoscopic conductor. We have presented
expressions for the relaxation rate, the decoherence rate
and the measurement rate in terms of capacitance
coefficients and in terms of potential derivatives
of the scattering matrix. The discussion illustrates
that scattering theory is useful not only for the discussion
of dc-transport processes in mesoscopic physics but in fact has
a much wider range of applicability. We have emphasized
screening in the detector:  as illustrated in Fig. \ref{Integrable QPC}
the decoherence rate and similarly the relaxation and measurement 
rate) are rappidly reduced with increasing channel number. 
Our discussion
is based only on a single fluctuating potential
in the detector but in principle a refined calculation which takes
into account a more realistic potential distribution
is possible \cite{AMMB}.

This work was supported by the Swiss National Science Foundation.\\
{\bf Note added in proof:}
Recently an instructive, information theoretical discussion of the
problem treated in this work was provided by Clerk,
Girvin, and Stone \cite{Clerk1}.


\begin{thebibliography}{150}


\bibitem{exp}

                       For recent experimental advances
                       in the read-out of dynamic states see:
                       D. Vion, A. Aassime, A. Cottet, P. Joyez,
                       H. Pothier, C. Urbina, D. Esteve,
                       and M. H. Devoret, Science {\bf 296}, 886 (2002).
                       Y. Yu, S. Han, X. Chu, S.-I. Chu, Z. Wang,
                       Science {\bf 296}, 889 (2002);
                       Y. Nakamura, Yu. A. Pashkin, and J. S. Tsai,
                       Nature (London) {\bf 398}, 357 (2001).

\bibitem{averin}       Exceptions are quantum non-demolition measurements.
                       For a recent proposal see:
                       D. V. Averin, Phys. Rev. Lett. {\bf 88},
                       207901 (2002).


\bibitem{Buks1}        E. Buks, R. Schuster, M. Heiblum, D. Mahalu, 
                       and V. Umansky, Nature {\bf 391}, 871 (1998).

\bibitem{Sprinzak1}    D. Sprinzak, E. Buks, M. Heiblum, and H. Shtrikman,
                       Phys. Rev. Lett. {\bf 84}, 5820 (2000).


\bibitem{Field1}        M. Field, C. G. Smith, M. Pepper,
                        D. A. Ritchie, J. E. F. Frost,
                        G. A. C. Jones, and D. G. Hasko,
                        Phys. Rev. Lett. {\bf 70}, 1311 (1993).


\bibitem{Smith1}       C. G. Smith, S. Gardelis, J. Cooper,
                       D. A. Ritchie, E. H. Linfield, Y. Jin,
                       H. Launois,
                       Physica E {\bf 12}, 830 (2002).

\bibitem{Levinson1}    Y. Levinson,
                       Europhys. Lett. {\bf 39}, 299  (1997).

\bibitem{Levit1}       A. Silva and S. Levit,
                       Phys. Rev. B {\bf 63}, 201309(R) (2001).

\bibitem{Aleiner1}     I. L. Aleiner, N. S. Wingreen, and Y. Meir,
                       Phys. Rev. Lett. {\bf 79}, 3740 (1997).

\bibitem{Stodolsky1}   R. A. Harris and L. Stodolsky,
                       Phys. Lett. B {\bf 116}, 464 (1982).

\bibitem{MBMartin1}    M. B\"uttiker and A. M. Martin,
                       Phys. Rev. B {\bf 61}, 2737 (2000).

\bibitem{Gurvitz1}     S. A. Gurvitz,
                       Phys. Rev. B {\bf 56}, 15215 (1997).

\bibitem{Schoen1}      Y. Makhlin, G. Sch\"on, and A. Schnirman,
                       Rev. Mod. Phys. {\bf 73}, 357 (2001).

\bibitem{Korotkov1}    A. N. Korotkov,
                       Phys. Rev. B {\bf 63}, 115403 (2001).

\bibitem{Milburn1}     H. S. Goan and G. J. Milburn,
                       Phys. Rev. B {\bf 64}, 235307 (2001).

\bibitem{Averin2}      D. V. Averin,
                       cond-mat/0010052 (2000).

\bibitem{Buettiker2}   M. B\"uttiker, H. Thomas, and A. Pr\^etre,
                       Phys. Lett. A {\bf 180}, 364 (1993).

\bibitem{Christen}     T. Christen and M. B\"uttiker,
                       Europhys. Lett. {\bf 35}, 523 (1996).

\bibitem{Ankara}       M. B\"{u}ttiker,
                       in "Quantum Mesoscopic Phenomena and Mesoscopic
                       Devices",
                       edited by I. O. Kulik and R. Ellialtioglu, (Kluwer,
                       Academic Publishers, Dordrecht,
                       2000). Vol. 559, p. 211.
                       cond-mat/9911188

\bibitem{Averin1}      A. N. Korotkov and D. Averin,
                       Phys. Rev. B {\bf 64}, 165310 (2001).

\bibitem{Pilgram1}     S. Pilgram and M. B\"uttiker,
                       Phys. Rev. Lett. {\bf 89},
                       200401 (2002).

\bibitem{Carmichael1}  H. J. Carmichael, {\it Statistical Methods in
                       Quantum Optics 1, 
                       Master Equations and Fokker-Planck
                       Equations}, (Springer, 
                       Berlin 1999).

\bibitem{Pedersen1}    M. H. Pedersen, S. A. van Langen, 
                       and M. B\"uttiker,
                       Phys. Rev. B {\bf 57}, 1838 (1998).

\bibitem{Landau1}      L. D. Landau and E. M. Lifshitz,
                       {\it Course of Theoretical Physics},
                       (Pergamon Press, Oxford 1977). 
                       Vol. {\bf 3}, \S 25. 
                       

\bibitem{AMMB}           A. M. Martin and M. B\"uttiker,
                         Phys. Rev. Lett. {\bf 84}, 3386 (2000).

\bibitem{Clerk1}         A. A. Clerk, S. M. Girvin, A. D. Stone, 
                         cond-mat/0211001 (2002).


\end{thebibliography}
\end{document}